\pdfoutput=1

      \documentclass[sort&compress,final]{aipproc}

      \layoutstyle{8x11single}

      \usepackage{graphicx}
      \usepackage{amsmath}
      \usepackage{amssymb}
      \usepackage{dsfont}
      \usepackage{wrapfig}

      \newcommand{\conjg}[1]{\ensuremath{\hspace{1pt}\overline{\hspace{-1pt}#1\hspace{-1pt}}}\hspace{1pt}}

      \def\Slash#1{\setbox0=\hbox{$#1$} 
      \dimen0=\wd0 
      \setbox1=\hbox{/} \dimen1=\wd1 
      \ifdim\dimen0>\dimen1 
      \rlap{\hbox to \dimen0{\hfil/\hfil}} 
      #1 
      \else 
      \rlap{\hbox to \dimen1{\hfil$#1$\hfil}} 
      / 
      \fi}

\begin{document}

       \title{Baryons in and beyond the quark-diquark model}


       \author{G. Eichmann}{
         address={Institut f\"{u}r Kernphysik, Technische Universit\"at Darmstadt, D-64289 Darmstadt, Germany}
       }

       \author{R. Alkofer}{
         address={Institut f\"ur Physik, Karl-Franzens-Universit\"at Graz, A-8010 Graz, Austria}
       }

       \author{C. S. Fischer}{
         address={Institut f\"{u}r Theoretische Physik, Universit\"at Giessen, D-35392 Giessen, Germany},
         altaddress={Institut f\"{u}r Kernphysik, Technische Universit\"at Darmstadt, D-64289 Darmstadt, Germany}
       }

       \author{A. Krassnigg}{
         address={Institut f\"ur Physik, Karl-Franzens-Universit\"at Graz, A-8010 Graz, Austria}
       }

       \author{D. Nicmorus}{
         address={FIAS, Johann Wolfgang Goethe-Universit\"{a}t, D-60438 Frankfurt am Main, Germany}
       }

       \begin{abstract}
        	We examine the nucleon's electromagnetic form factors in a Poincar\'{e}-covariant Faddeev framework.
        	The three-quark core contributions to the form factors are obtained by employing a quark-diquark approximation.
            We implement the self-consistent solution for the quark-photon vertex from its inhomogeneous Bethe-Salpeter equation.
            We find that the resulting transverse parts which add to the Ball-Chiu vertex have no significant impact on nucleon magnetic moments.
            The current-quark mass evolution of the form factors agrees with results from lattice QCD.
       \end{abstract}

       \maketitle


         \section{Introduction}

              Investigating the nucleon's electromagnetic structure continues
              to be an experimental challenge in contemporary particle physics.
              High-precision electron scattering measurements have provided information on the nucleon's
              electromagnetic form factors as well as  
              longitudinal vs. transverse momentum fractions and  
              the distribution of spin and orbital angular momentum among its constituents;
              see~\cite{Arrington:2006zm,Perdrisat:2006hj} for reviews.
              Approaching these issues from a theoretical perspective aims
              at an understanding of the hadrons' substructure in terms quarks and gluons,  
              the fundamental degrees of freedom in Quantum Chromodynamics.

              In a Dyson-Schwinger/bound-state equation approach (see~\cite{Alkofer:2000wg,Fischer:2006ub,Roberts:2007jh} for reviews)
              baryons are obtained as solutions of the covariant Faddeev equation.
              It describes the binding of a baryon by iterated quark-quark correlations and thereby constitutes
              the three-body analogue of a quark-antiquark Bethe-Salpeter equation (BSE).
              The nucleon's Faddeev equation was recently solved in a numerical setup where the full covariant structure
              of the nucleon amplitude was implemented~\cite{Eichmann:2009qa}. The quark-quark kernel was modeled by a rainbow-ladder interaction,
              i.e. a dressed gluon exchange between any two quarks, where the quark propagator was consistently determined from its Dyson-Schwinger equation (DSE).
              A rainbow-ladder truncation excludes the presence of pion-cloud corrections which have significant impact on the chiral structure of hadrons.
              It therefore reflects the properties of a hadronic 'quark core'.

              The computation of electromagnetic form factors in this setup requires knowledge of another ingredient,
              namely the dressed quark-photon vertex.
              A frequently used truncation of the Faddeev equation is the quark-diquark model, where
              the binding of baryons is traced back to the interaction between quarks and effective scalar and axial-vector diquarks
              within a baryon~\cite{Hellstern:1997pg,Eichmann:2007nn}. It was demonstrated in Ref.\,\cite{Eichmann:2009qa} that this simplification only introduces
              a $\sim 5\%$ change in the nucleon mass. Hence, as an intermediate step towards a fully self-consistent
              determination of electromagnetic form factors from the covariant Faddeev equation, we investigate
              the impact of the structure of the quark-photon vertex on nucleon form factors in the quark-diquark approach.

         \section{Nucleon electromagnetic form factors}

              The nucleon's electromagnetic current is expressed in terms of two form factors: the Dirac and Pauli form factors $F_1(Q^2)$ and $F_2(Q^2)$,
              or the Sachs form factors as their linear combinations $G_E = F_1 - \tau\,F_2$ and $G_M = F_1+F_2$, with $\tau = Q^2/(4M_N^2)$.
              Their static values represent the proton and neutron charges $\lambda^{p,n}=(1,0)$, (anomalous) magnetic moments $\kappa^{p,n}$ and $\mu^{p,n}$,
               and electromagnetic radii $r_{1,2}^{p,n}$ and $r_{E,M}^{p,n}$:
    \renewcommand{\arraystretch}{1.2}
              \begin{equation}
                 \begin{array}{rl}
                   F_1(0) \!\!\!\!\!\!&= \lambda\,, \\
                   G_E(0) \!\!\!\!\!\!&= \lambda\,,
                 \end{array} \quad
                 \begin{array}{rl}
                   F_2(0) \!\!\!\!\!\!&= \kappa\,, \\
                   G_M(0) \!\!\!\!\!\!&= \mu\,,
                 \end{array} \quad
                 \begin{array}{rl}
                   r_1^2 \!\!\!\!\!\!&= -6 \,F'_1(0)\,, \\
                   r_E^2 \!\!\!\!\!\!&= -6 \,G'_E(0)\,,
                 \end{array} \quad
                 \begin{array}{rl}
                   \kappa \,r_2^2 \!\!\!\!\!\!&= -6 \,F'_2(0)\,, \\
                      \mu \,r_M^2 \!\!\!\!\!\!&= -6 \,G'_M(0)\,.
                 \end{array}
              \end{equation}

              Electromagnetic current conservation provides a construction principle for the current in a given model framework~\cite{Oettel:1999gc}.
              In the quark-diquark approach this construction involves quark and diquark impulse-approximation diagrams, but also
              a coupling to the exchanged quark between quark and diquark as well as seagull terms, i.e. the photon's coupling to the diquark amplitudes.
              Upon resolving the diquark-photon vertex into its constituents, all photon couplings are specified by two quantities:
              the quark-photon vertex and the seagull vertices.
              The details of this construction are described in Refs.\,\cite{Eichmann:2007nn,Eichmann:2009zx} for the nucleon
              and Ref.\,\cite{Nicmorus:2010sd} for the $\Delta$-baryon, respectively.

             Herein we want to investigate the structure of the quark-photon vertex in the context of electromagnetic form factors.
             The most general form of the vertex can be expressed by a sum of the Ball-Chiu term~\cite{Ball:1980ay} and a purely transverse contribution:
             \begin{equation}\label{qpv-general}
                \Gamma^\mu_\text{q}(k,Q) = \Big[i\gamma^\mu \Sigma_A + 2 k^\mu \left(i\Slash{k}\,\Delta_A + \Delta_B\right)\Big] +\Gamma^\mu_T(k,Q)\,.
             \end{equation}
             The Ball-Chiu part involves the quantities
             \begin{equation}\label{QPV:sigma,delta}
                \Sigma_A = \frac{A(k_+^2)+A(k_-^2)}{2} , \quad
                \Delta_A = \frac{A(k_+^2)-A(k_-^2)}{k_+^2-k_-^2}, \quad
                \Delta_B = \frac{B(k_+^2)-B(k_-^2)}{k_+^2-k_-^2},
             \end{equation}
             where $k_\pm = k \pm Q/2$ are the in- and outgoing quark momenta and
             $A(k^2)$ and $B(k^2)=M(k^2)A(k^2)$ are the dressing functions of the quark propagator
             $S(k)=A^{-1}(-i \Slash{k}+M)/(k^2 +M^2)$.
             Eq.\,\eqref{qpv-general} satisfies the Ward-Takahashi identity $Q^\mu \Gamma^\mu_\text{q}(k,Q) = S^{-1}(k_+)-S^{-1}(k_-)$.
             The transverse part can be written as
             \begin{equation}\label{transverse-vertex}
                -i \Gamma^\mu_T = f_1\,\gamma^\mu_T  + if_2\, \gamma^\mu_T\Slash{Q}
                                  + \textstyle\frac{i}{2}\, f_3\, k\! \cdot \! Q  \left[\gamma^\mu_T, \,\Slash{k} \right] 
                                  + f_4\,\gamma^\mu_t \Slash{k}_T \Slash{Q}
                                    + k^\mu_T\big(i f_5 + f_6\,k\! \cdot \! Q\,\,\Slash{Q}  + f_7\,\Slash{k}  + i f_8 \,\Slash{k}_T \Slash{Q} \big)\,,
             \end{equation}
             where the $f_i(k^2, \,k\cdot Q, \,Q^2)$ are scalar dressing functions.
             We used the abbreviations $\gamma^\mu_T = T^{\mu\nu}_Q \gamma^\nu$, $k^\mu_T = T^{\mu\nu}_Q k^\nu$ and $\gamma^\mu_t = T^{\mu\nu}_{k_T} \gamma^\nu_T$,
             with $T^{\mu\nu}_Q = \delta^{\mu\nu} - \hat{Q}^\mu \hat{Q}^\nu$ being a transverse projector with respect to the photon momentum $Q$.

             The quark-photon vertex is obtained from its inhomogeneous Bethe-Salpeter equation~\cite{Maris:1999bh}, given by
             \begin{equation}\label{qpv-bse}
                  \left[\Gamma^\mu_\text{q}(k,Q)-Z_2\,i\gamma^\mu\right]_{\alpha\beta} =
                  \frac{4}{3} \int_{p'} K_{\alpha\alpha'\beta'\beta}\left[ S(k_+')\,\Gamma^\mu_\text{q}(k',Q)\,S(k_-')\right]_{\alpha'\beta'}\,,
             \end{equation}
             where $Z_2$ is the quark renormalization constant and $K_{\alpha\alpha'\beta'\beta}$ the rainbow-ladder kernel
             that also appears in the quark DSE and the homogeneous meson and diquark BSEs. It reads
            \begin{equation}\label{RLkernel}
                K_{\alpha\alpha'\beta\beta'} =  Z_2^2 \, \frac{ 4\pi \alpha(k^2)}{k^2} \, T^{\mu\nu}_k \gamma^\mu_{\alpha\alpha'} \,\gamma^\nu_{\beta\beta'}\,,
            \end{equation}
            with $k$ being the gluon momentum,
            and involves an effective interaction $\alpha(k^2)$ which defines the model input of the approach.
            We use the parametrization of Ref.\,\cite{Maris:1999nt} for $\alpha(k^2)$ 
            whose infrared term is modeled by two parameters: an infrared scale $\Lambda$ and a width parameter $\eta$.
            Adjusting the scale $\Lambda$ to reproduce the experimental pion decay constant yields a good description of pseudoscalar-meson,
            vector-meson, nucleon and $\Delta$ ground states. Inflating this scale towards the chiral limit mimicks the properties of an effective quark core:
            it generates hadron masses which are overestimated in the chiral region and
             must be dressed, for instance, by pionic corrections; see Refs.\,\cite{Eichmann:2008ae,Nicmorus:2010sd} for a further discussion.
            These studies exhibited only a modest dependence on the width parameter $\eta$; i.e., the quoted observables
            are not sensitive to the details of the interaction in the infrared.

             To solve the inhomogeneous BSE it is convenient to express the vertex in an orthonormal basis, for instance defined by the elements
             $\tau_i^\mu(k,Q) \in\{ \gamma^\mu_t\!/\!\sqrt{2}, \;\widehat{k_T}^\mu, \; \widehat{Q}^\mu \} \times
             \{ \mathds{1}\,,\; \widehat{\Slash{Q}}\,,\; \widehat{\Slash{k}_T}\,,\; \widehat{\Slash{k}_T}\,\widehat{\Slash{Q}} \}$
             which satisfy the orthogonality relation $\text{Tr} \{ \conjg{\tau}_i^\mu \tau_j^\mu  \} = 4\delta_{ij}$.
             Since the inhomogeneous equations for the transverse and longitudinal parts decouple,
             it is sufficient to consider the eight transverse elements alone. 
             The longitudinal result reproduces the longitudinal projection
             of the Ball-Chiu vertex and thereby the Ward-Takahashi identity,
             and purely longitudinal terms do not contribute to nucleon form factors because of current conservation: $Q^\mu J^\mu = 0$.
             The inhomogeneous BSE self-consistently generates a timelike vector-meson pole in the quark-photon vertex
             at $Q^2 =-m_\rho^2$ which significantly increases the charge radii of pseudoscalar and vector mesons~\cite{Maris:1999bh,Bhagwat:2006pu}.

    \renewcommand{\arraystretch}{1.0}

             \begin{table}[t]

                \begin{tabular}{   l @{\;\;}  || @{\;\;}c@{\;\;} | @{\;\;}c@{\;\;}  ||  @{\;\;}c@{\;\;} | @{\;\;}c@{\;\;} | @{\;\;}c@{\;\;} | @{\;\;}c@{\;\;}   }

                                       &  $\mu^p$      &  $\mu^n$     &  $R_E^p$   &  $R_E^n$   &  $R_M^p$  &  $R_M^n$   \\   \hline

                    BC + $\rho$        &  $2.56(6)$    &  $-1.59(3)$  &  $0.55(2)$ &  $0.00(1)$ &  $0.46(1)$ &  $0.45(1)$             \\
                    Full vertex        &  $2.52(12)$   &  $-1.59(7)$  &  $0.58(1)$ &  $0.01(1)$ &  $0.50(1)$ &  $0.49(1)$             \\ \hline
                    Exp.               &  $2.79$       &  $-1.91$     &  $0.69(2)$    &  $-0.10$   &  $0.64(5)$ &  $0.67(2)$

                \end{tabular} \caption{Static electromagnetic properties of the nucleon at the physical pion mass $m_\pi=140$ MeV.
                                       The magnetic moments $\mu^{p,n}$ are dimensionless and the squared charge radii $R_{E,M}^{p,n} = (r_{E,M}^{p,n} \,M_N)^2$
                                       are given in (GeV\,fm)$^2$.
                                       The first row shows the results obtained with a Ball-Chiu vertex augmented by the ansatz~\eqref{rho-ansatz};
                                       the second row those with the quark-photon vertex solution from the inhomogeneous BSE.
                                       The brackets indicate the sensitivity to the model parameter $\eta$.
                                       The last row collects experimental results~\cite{Belushkin:2006qa,Nakamura:2010zzi}.
                                       }\label{tab:results}

        \end{table}

         \section{Results and discussion}

             Upon implementing the numerical solution for the quark-photon vertex according to Eqs.\,(\ref{qpv-bse}--\ref{RLkernel})
             in the nucleon form factor setup,
             one can quantify the impact of the purely transverse structure \eqref{transverse-vertex} of the vertex.
             In previous form factor studies this transverse part was modeled by an effective $\rho-$meson contribution~\cite{Eichmann:2008ef,Eichmann:2009zx}:
             \begin{equation}\label{rho-ansatz}
                \Gamma^\mu_T = -\frac{1}{g_\rho}\,\frac{Q^2}{Q^2+m_\rho^2}\,e^{-g(Q^2/m_\rho^2)}\,\Gamma_\text{vc}(k,Q)\,,
             \end{equation}
             where $\Gamma_\text{vc}(k,Q)$ is the calculated vector-meson amplitude (see Ref.\,\cite{Eichmann:2008ef} for details).
             In consistency with the behavior of the pion's charge radius, 
             the Ball-Chiu part and Eq.\,\eqref{rho-ansatz}
             each roughly contribute $\sim 50\%$ to the nucleon's squared charge radii
             $(r_E^p)^2$ and $(r_M^{p,n})^2$, whereas Eq.\,\eqref{rho-ansatz}
             cancels the Ball-Chiu piece in the neutron's electric charge radius $(r_E^n)^2$.
             On the other hand, this ansatz does not contribute to the nucleon magnetic moments due to the inherent suppression with~$Q^2$.
             The resulting static electromagnetic properties allow for the interpretation as a quark core which agrees well
             with lattice data at larger quark masses where meson-cloud effects are no longer important.

             It is visible from Table\,\ref{tab:results} that the full vertex-BSE solution does not significantly change this behavior.
             The resulting nucleon charge radii differ by $\lesssim 5\%$ from those obtained via Eq.\,\eqref{rho-ansatz}.
             The fractions of $\mu^p$ and $\mu^n$ which come from the transverse vertex-BSE solution of Eq.\,\eqref{transverse-vertex}
             owe to the amplitudes $f_2$ and $f_4$. Those two substantially contribute to a quark's anomalous magnetic moment~\cite{Chang:2010hb}
             where they roughly cancel the Ball-Chiu part.
             Nonetheless, they have only a small impact upon the nucleon magnetic moments which are still sufficiently well
             represented by the Ball-Chiu ansatz alone.
             Moreover, we have verified that the $Q^2-$evolution of all electromagnetic form factors is essentially identical
             for both choices of transverse contributions up to $Q^2 \sim 3$ GeV$^2$.

\begin{figure}[htp]
               \includegraphics[height=.43\textheight]{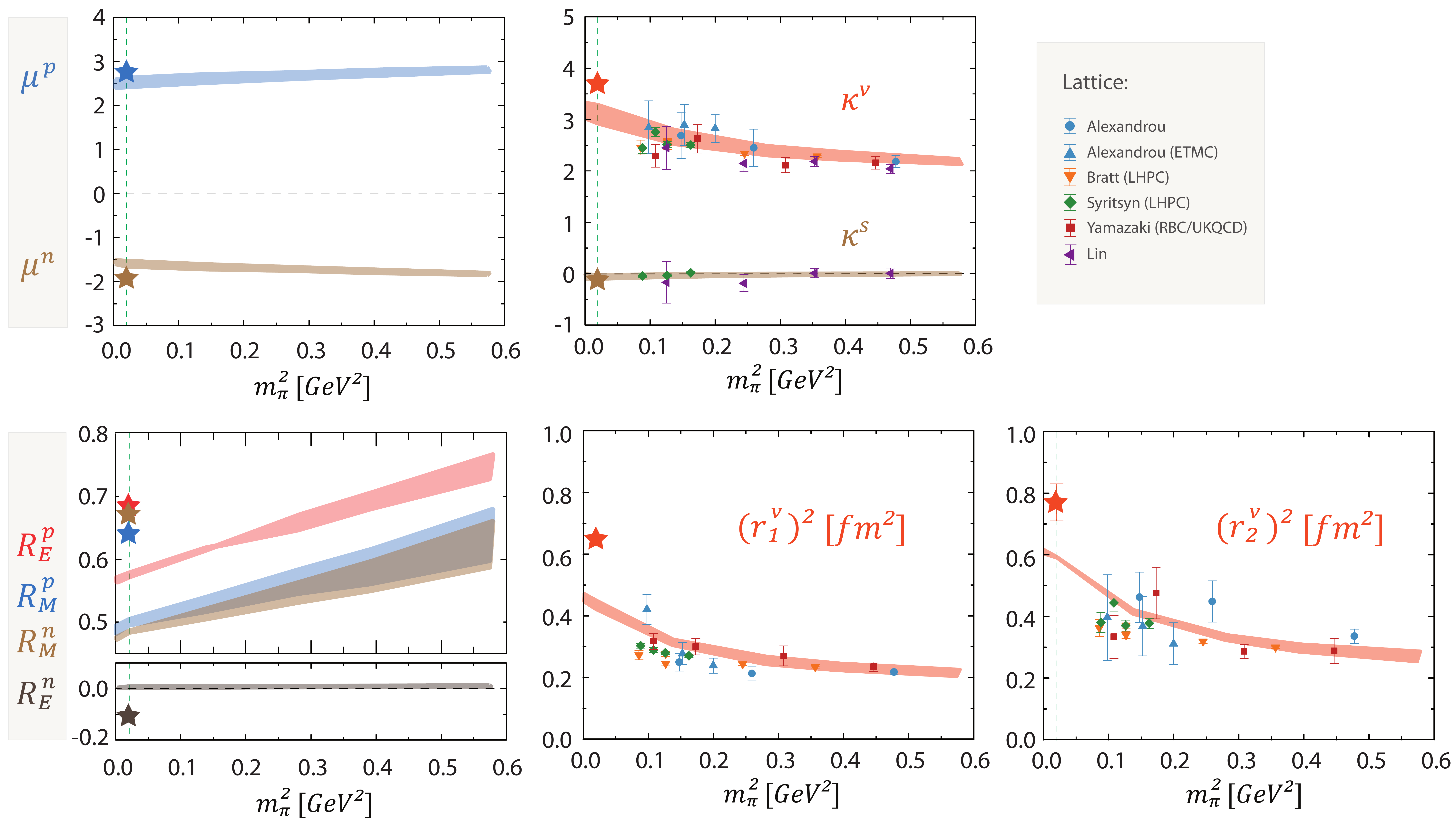}
               \caption{Static electromagnetic properties of the nucleon as functions of the squared pion mass.
                        \emph{Left panels:} dimensionless magnetic moments and squared charge radii in units of (GeV\,fm)$^2$.
                        \emph{Middle and right panels:} isovector and isoscalar anomalous magnetic moments and squared isovector charge radii,
                        rescaled as described in the text and compared to a selection of recent
                        lattice results~\cite{Alexandrou:2006ru,Alexandrou:2008rp,Syritsyn:2009mx,Yamazaki:2009zq,Bratt:2010jn,Lin:2010fv}.
                        The bands show the dependence on the model parameter $\eta$; stars denote experimental values~\cite{Belushkin:2006qa,Nakamura:2010zzi}.   }
               \label{fig}
             \end{figure}

             In Fig.\,\ref{fig} we show the current-mass evolution of the nucleon magnetic moments and charge radii.
             The left panels depict the dimensionless magnetic moments $\mu^{p,n}$ and squared charge radii $R_{E,M}^{p,n} := (r_{E,M}^{p,n}\,M_N)^2$
             which exhibit a moderately rising behavior with the pion mass.
             In the remaining panels we show the isovector and isoscalar anomalous magnetic moments and isovector Dirac and Pauli radii
             as obtained from the form factor combinations $F_i^v = F_i^p-F_i^n$ and $F_i^s = F_i^p+F_i^n$, with $i=1,2$.
             To allow for a sensible comparison of dimensionful charge radii and magnetic moments in
             static nuclear magnetons with lattice results  
             we plot the following rescaled quantities:
             \begin{equation}\label{magneticmoment}
                 \kappa_\text{resc} = \kappa\,\frac{M_N^\text{exp}}{M_N^\text{Ref}}\,, \quad
                 r^2_\text{resc} = r^2\left(\frac{M_N}{M_N^\text{Ref}}\right)^2\,, \quad \text{with} \quad
                 M_N^\text{Ref}(m_\pi^2)^2 = M_0^2 + \left(\frac{3 m_\pi}{2}\right)^2 \left(1 + f(m_\pi^2)\right) ,
             \end{equation}
             where $\kappa$, $r$ and $M_N$ are our calculated results and
             $M_N^\text{Ref}$, with $f(m_\pi^2) = 0.77/(1+(m_\pi/0.65\,\text{GeV})^4)$ and $M_0=0.9$~GeV,
             is a reference mass which 
             describes the lattice results for $M_N$ of Refs.\,\cite{Alexandrou:2006ru,Alexandrou:2008rp,Syritsyn:2009mx,Yamazaki:2009zq,Bratt:2010jn,Lin:2010fv} reasonably well.

             Fig.\,\ref{fig} shows a good agreement between our calculated magnetic moments and the lattice data.
             This is especially relevant at higher pion masses where the pion-cloud dressing effects of the nucleon are diminished. In combination with
             the observed missing chiral curvature, our results can be viewed to represent the quark core of the nucleon. The implementation
             of such chiral effects, together with a form factor determination beyond the quark-diquark model, remains a task for future investigations.

       \begin{theacknowledgments}
         We are grateful to M. Blank, I. C. Clo\"et, and R. Williams for discussions.
         This work was supported by the Austrian Science Fund FWF under
          Projects No.~J3039-N16, P20592-N16 and P20496-N16;
          by the Helmholtz Young Investigator Grant VH-NG-332;
          by HIC for FAIR within the LOEWE program launched by the State of Hesse, GSI, BMBF and DESY;
          and in part by the European Union (HadronPhysics2 project "Study of strongly interacting matter").
       \end{theacknowledgments}

\newpage

       \bibliographystyle{aipproc-mod}    

       \bibliography{lit}

\end{document}